\documentclass[preprint]{jpsj3}

\usepackage{graphicx}

\title{Microscopic Analysis of Resonant Inelastic X-Ray Scattering in Orbital-Ordered KCuF$_3$}

\author{Takuji \textsc{Nomura}\thanks{E-mail address: nomurat@spring8.or.jp} }

\inst{
Quantum Beam Science Center, Japan Atomic Energy Agency, 
Sayo, Hyogo 679-5148, Japan \\}

\recdate{\today}

\abst{
We analyze resonant inelastic x-ray scattering (RIXS) at the Cu $K$ edge 
in a typical orbital-ordered compound KCuF$_3$ on the basis of a microscopic theory. 
Spectral shape and its dependence on polarization direction and momentum transfer 
of photons are explained consistently with experimental data within our microscopic calculation. 
According to our microscopic orbital-resolving analysis, 
high-energy spectral weights (above 5 eV) originate from charge-transfer excitations 
related to the Cu-$d\gamma$ orbitals, while the low-energy weights (below 2 eV) originate 
from the $d$-$d$ orbital excitations among the five Cu-$d$ orbitals. 
We assign specifically the RIXS weights to microscopic orbital-excitation processes, 
beyond the previous phenomenological assignment based on symmetry properties. 
}

\begin{document}
\maketitle

\section{Introduction}

Resonant inelastic x-ray scattering (RIXS) is growing up to be a powerful method 
of measuring elementary excitations in solids~\cite{Ament2011}. 
Among RIXS phenomena, RIXS at the transition-metal $K$ edges attracts much interest, 
because it provides to us a unique technique to observe charge and orbital excitations 
in strongly correlated $d$ electrons of transition-metal compounds
~\cite{Hasan2000,Kim2002,Ishii2005,Suga2005,Ishii2011,Jarrige2012}. 
Here we illustrate the RIXS process at the transition-metal $K$ edge: 
firstly, an incident photon with the energy tuned to the transition-metal $K$ edge 
is resonantly absorbed to promote an inner-shell $1s$ electron to the $4p$ conduction bands 
(in the case of 3$d$ transition-metal compounds), 
following the dipole-transition rule, and in the intermediate state 
a hole is created at the local $1s$ orbital. 
The created $1s$ hole plays a role of a local scattering body for the electrons near the Fermi level. 
In other words, electrons near the Fermi level are excited to screen the created $1s$ hole. 
Before the excitation near the Fermi level damps, the initially excited $4p$ electron goes back 
to the $1s$ state to fill the $1s$ hole, with emitting a photon. 
Following the energy-momentum conservation law, the emitted photon should 
have energy and momentum which differ from those of the incident photon 
by the amount of energy and momentum spent to excite the electrons near the Fermi level. 
Here we should note that not all of the electrons near the Fermi level are evenly excited 
in the intermediate and final states: electrons only weakly interacting with the $1s$ hole 
are not strongly excited. 
On the other hand, electrons strongly interacting with the $1s$ hole can be strongly excited. 
Transition-metal $d$ electrons ($3d$ electrons in the cases of 3$d$ transition-metal compounds) 
are relatively localized in space, and therefore the Coulomb interaction 
between the $1s$ and $d$ orbitals is expected to be strong. 
Thus, RIXS at the $K$ edge in transition-metal compounds enables us to observe 
selectively the excitations of strongly correlated $d$ electrons. 

RIXS in transition-metal compounds has been studied intensively also from theoretical sides. 
This is partly because RIXS is a relatively difficult phenomenon to interpret 
without intricate theoretical considerations. 
RIXS not only involves complex intermediate excitation processes 
but is subject to strong electron correlations, as easily understood. 
Thus, RIXS has provided good opportunities of applying various 
theoretical approaches, e.g., numerical diagonalization for finite-size clusters\cite{Tsutsui1999}, 
many-body perturbation theory~\cite{Nomura2004,Nomura2005}, 
dynamical mean-field theory~\cite{Pakhira2012} and so on~\cite{Ide1999,Brink2006}. 
Among several theoretical studies, we previously derived a useful formula 
to calculate RIXS spectra by taking account of the above mentioned RIXS process 
within the Keldysh perturbative formalism~\cite{Nomura2004,Nomura2005,Igarashi2006}, 
and analyzed RIXS spectral properties for many transition-metal 
compounds~\cite{Nomura2004,Nomura2005,Igarashi2006,Takahashi2007,Semba2008,Nomura2012}. 
In our previous works~\cite{Nomura2004,Nomura2005,Igarashi2006} 
and most of others' works, the excited $4p$ electron 
has been considered to be a `spectator'~\cite{Kim2007}. 
In fact, the electronic structure for the $4p$ bands 
has been highly simplified or treated only crudely in most of previous works
~\cite{Ide1999,Tsutsui1999,Nomura2004,Nomura2005,Brink2006}. 

Recently, Ishii and collaborators measured RIXS at the Cu $K$ edge 
in a typical orbital-ordered compound KCuF$_3$~\cite{Ishii2011}. 
They observed that the low-energy RIXS weights show notable characteristic dependence 
on polarization direction of photons, while they did not observed any notable momentum dependence. 
They attributed the observed low-energy features to possible $d$-$d$ excitation processes
phenomenologically on the basis of symmetry properties. 
According to the dipole-transition rule, the polarization direction is closely related 
to the excited $4p$ state in the intermediate state of RIXS. 
Therefore, such notable polarization dependence suggests strongly 
that the $4p$ electron plays a much more important role than a `spectator'. 

The aim of our present study is to analyze theoretically spectral shape and its dependence 
on the polarization direction and momentum transfer of photons within a microscopic calculation. 
In addition, we elucidate microscopically relevant orbital-excitation processes in RIXS of KCuF$_3$, 
by introducing our new method of orbital-resolving analysis. 
The present article is constructed as follows: 
In \S~\ref{Sc:Formulation}, we present our Hamiltonian and perturbative formulation for RIXS intensity. 
We also define orbital-resolved RIXS spectra. 
To describe the electronic structure of KCuF$_3$ precisely, 
we use first-principles band structure calculation, and determine the antiferromagnetic ground state 
within the Hartree-Fock (HF) approximation. 
Electron correlations in the intermediate states are treated within the random-phase approximation (RPA). 
In \S~\ref{Sc:Results}, we present numerical results on RIXS spectra and their dependences 
on polarization and momentum transfer of photons. 
The origin of each spectral weight is microscopically analyzed in detail 
by resolving orbital-excitation processes. 
In \S~\ref{Sc:Discussions}, some discussions and remarks on our formulation and results 
are given. In \S~\ref{Sc:Conclusions}, the article is concluded with summary. 

\section{Formulation of RIXS}
\label{Sc:Formulation}

To discuss the RIXS process microscopically, 
we consider the following form of Hamiltonian: 
\begin{equation} 
H = H_{n.f.} + H_{1s} + H_{1s-d} + H_x, 
\end{equation}
where $H_{1s}$ and $H_x$ describe the inner-shell $1s$ electrons 
and the dipole-transition by x-rays, respectively. 
$H_{n.f.}$ describes the correlated electrons near the Fermi level. 
$H_{1s-d}$ is the Coulomb interaction between $1s$ and transition-metal $d$ electrons. 
For $1s$ electrons, we take completely localized $1s$ orbitals at each transition-metal site: 
\begin{equation}
H_{1s} = \sum_i^{\rm t.m.} \sum_{\sigma} \varepsilon_{1s}(\mib{r}_i) s_{i\sigma}^{\dag} s_{i\sigma} 
= \sum_{\mib{k} \sigma} \varepsilon_{1s} s_{{\mib k}\sigma}^{\dag} s_{{\mib k}\sigma}, 
\end{equation}
where $\varepsilon_{1s}(\mib{r}_i) \equiv \varepsilon_{1s}$ is the one-particle energy 
of the $1s$ state, $s_{i\sigma}^{\dag}$ and $s_{i\sigma}$ are the creation 
and annihilation operators of $1s$ electrons with spin $\sigma$ 
at transition-metal site $i$, respectively. 
`$\rm{t.m.}$' in the summation with respect to $i$ means summing only 
over transition-metal sites. 
$ s_{{\mib k}\sigma} (s_{{\mib k}\sigma}^{\dag})$ is the momentum representation 
of $ s_{i \sigma} (s_{i \sigma}^{\dag})$. 
$H_x$ describes resonant $1s$-$4p$ dipole transition induced by x-rays: 
\begin{equation}
H_x = \sum_{\mib{k},\mib{q}} \sum_{\mu}^{xyz} \sum_{\sigma} 
w_{\mu}({\mib q},{\mib e}) \alpha_{\mib{q}\mib{e}} 
p_{{\mib k}+{\mib q} \mu \sigma}^{\dag} s_{{\mib k}\sigma} + h.c., 
\end{equation} 
where $p_{{\mib k} \mu \sigma}^{\dag}$ is the creation operator 
of transition-metal $4p_{\mu}$ electron ($\mu=x,y,z$), and 
$\alpha_{\mib{q}\mib{e}}$ is the annihilation operator of a photon 
with momentum $\mib{q}$ and polarization $\mib{e}$. 
The summation in $\mu$ with `$xyz$' at the top means that $\mu$ 
takes $x$, $y$ or $z$. 
We assume the matrix elements of  $w_{\mu}(\mib{q}, \mib{e})$ are given in the form: 
\begin{equation}
w_{\mu}(\mib{q}, \mib{e}) = - \frac{e}{m}\sqrt{\frac{2\pi}{|\mib{q}|}} \mib{e} 
\cdot \langle 4p_{\mu} |\mib{p}|1s \rangle \propto \mib{e} \cdot \mib{e}_\mu, 
\end{equation}
in natural units ($c = \hbar = 1$). 
$\mib{e}_\mu$'s are the orthonormal basis vectors. 
$H_{1s-d}$ is given by
\begin{equation}
H_{1s-d} = \sum_i^{\rm t.m.} \sum_{\sigma\sigma'} V_{1s-d}(\mib{r}_i) 
s_{i\sigma}^{\dag} d_{i\sigma'}^{\dag} d_{i\sigma'} s_{i\sigma}, 
\end{equation}
where $V_{1s-d}(\mib{r}_i)$ is the so-called core-hole potential 
at transition-metal site $\mib{r}_i$. 
In the present study on KCuF$_3$, we take $V_{1s-d}(\mib{r}_i) \equiv V_{1s-d} = 9$ eV 
at Cu sites.   

\begin{figure}
\begin{center}
\includegraphics[width=90mm]{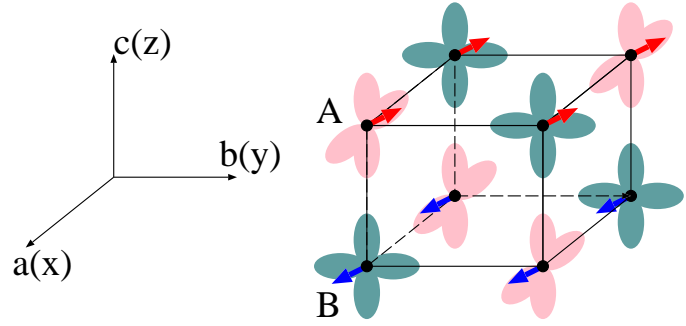}
\end{center}
\caption{
(Color online)
Schematic representation of the orbital-ordered state 
of KCuF$_3$~\cite{Kadota1967,Kugel1973,Towler1995,
Liechtenstein1995,Pavarini2008}. 
Cu atoms are placed on the corners of the pseudo-cubic cell 
(K and F sites are not shown explicitly). 
There are two kinds of Cu sites, depending on the orbital state: 
At A sites, Cu-$d_{x^2-z^2}$ states are filled almost by one half with electrons, 
while at B sites Cu-$d_{y^2-z^2}$ states are. 
Thick arrows represent the direction of the spin moment at each Cu site. 
The spin moments are parallel along the $ab$ plane 
in the antiferromagnetic ground state.}
\label{Fig:lattice}
\end{figure}

To prepare the Hamiltonian part $H_{n.f.}$, firstly we perform first-principles 
band structure calculation assuming the paramagnetic state~\cite{Blaha2012}. 
The orbital-ordered state is schematically represented in Fig.~\ref{Fig:lattice}. 
To express the electronic orbital bases and scattering geometry, 
we throughout take the coordinate system where the principal axes 
of the pseudo-cube constructed by Cu sites are parallel 
along the cartesian axes (see Fig~\ref{Fig:lattice}). 
We assume the so-called `a-type' structure 
(space group: $D_{4h}^{18}-I4/mcm$)~\cite{Okazaki1969,Hutchings1969}, 
and use the structure parameters given in Ref.~\ref{Okazaki1961}. 
Then we perform tight-binding fitting to the obtained energy bands 
near the Fermi level by using the {\tt wannier90} code~\cite{Mostofi2008, Kunes2010}, 
where we take $p$ and $d$ orbitals at K sites, $s$, $p$ and $d$ orbitals at Cu sites, 
and $p$ orbitals at F sites. Here we should interpret these $s$ orbitals  
at Cu sites as $4s$ orbitals, and not confuse with the $1s$ orbitals. 
Thus we include 52 localized Wannier states in the unit cell, 
because there are two K, two Cu and six F sites in the unit cell. 
Concerning the $d$ orbitals at Cu sites, 
we take $d(xy,yz,xz,x^2-z^2,3y^2-r^2)$ at A sites 
and $d(xy,yz,xz,y^2-z^2,3x^2-r^2)$ at B sites, 
where the orbital bases are defined following the coordinate axes in Fig.~\ref{Fig:lattice}. 
Thus we obtain a tight-binding model to fit the 52 bands in the energy window 
from $-8$ eV to 20 eV with respect to the Fermi energy. 
The reason why we choose those 52 localized orbitals is that those orbitals 
occupy the main part of the density of states in this energy window, 
according to the band structure calculation. 
We take about 14000 hoppings $t_{\ell\ell'}({\mib r})$ 
with ${\mib r}$ up to at most 10 lattice units. 
Adding the on-site Coulomb interaction part, 
we have the Hamiltonian part $H_{n.f.}$ in the following form: 
\begin{equation}
H_{n.f.} = \sum_{ii'} \sum_{\ell\ell'} \sum_\sigma t_{\ell\ell'}({\mib r}_i-\mib{r}_{i'}) 
a_{i \ell \sigma}^{\dag} a_{i' \ell' \sigma} + 
\frac{1}{2} \sum_i^{\rm t.m.} \sum_{\ell_{1 \sim 4}}^{@\mib{r}_i} 
\sum_{\sigma\sigma'} I_{\ell_1\ell_2;\ell_3\ell_4}(\mib{r}_i) 
a_{i \ell_1 \sigma}^{\dag} a_{i \ell_2 \sigma'}^{\dag} 
a_{i \ell_3 \sigma'} a_{i \ell_4 \sigma}, 
\end{equation}
where $a_{i \ell \sigma}^{\dag}$ and $a_{i \ell \sigma}$ 
are the electron creation and annihilation operators 
for orbital $\ell$ with spin $\sigma$ at site $i$. 
$I_{\ell_1\ell_2;\ell_3\ell_4}(\mib{r}_i) \equiv I_{\ell_1\ell_2;\ell_3\ell_4}$ 
is the on-site Coulomb integral at transition-metal (i.e., Cu) sites. 
In the summation with respect to $\ell_n$, `$@\mib{r}_i$' at the top 
means orbital $\ell_n$ should be placed on the site $\mib{r}_i$. 
One-particle energy at orbital $\ell$ is given by $\varepsilon_{\ell} \equiv t_{\ell\ell}(\mib{r}=0)$. 
We modify the one-particle energy $\varepsilon_{\ell}$ for Cu-$d$ orbitals, to obtain 
a realistic level scheme of the local Cu-$d$ orbitals, as explained in Appendix. 
Hereafter we use the following convention: if $\ell$ denotes $d$ orbital (e.g., $\ell=xy$), 
then $a_{i \ell \sigma} \equiv d_{i \ell \sigma}$, 
if $\ell$ denotes $p$ orbital (e.g., $\ell=x$), 
then $a_{i \ell \sigma} \equiv p_{i \ell \sigma}$, and so on. 
$a_{i \ell \sigma}$ contains also the annihilation operators at K and F sites. 
However, we expect that the above convention does not cause 
any confusion among the operators for K-$p$, Cu-$p$ and F-$p$ orbitals, 
or between the operators for K-$d$ and Cu-$d$ orbitals, 
because the operators for orbitals at K and F sites do not appear explicitly in the present article. 
Here we introduce the values of on-site Coulomb interaction $I_{\ell_1\ell_2;\ell_3\ell_4}$ 
at each Cu site in the form of Slater-Condon integrals 
(see Ref.~\ref{Condon1959} for the definition of Slater-Condon integrals and their relation 
to $I_{\ell_1\ell_2;\ell_3\ell_4}$): $F_{dd}^0 = 10.5$ eV, 
$F_{dd}^2 = 12$ eV, $F_{dd}^4 = 8$ eV. 
These values of $F_{dd}^2$ and $F_{dd}^4$ are similar to those determined 
for copper oxides in Ref.~\ref{Czyzyk1994} (11.5 eV and 7.4 eV, respectively, there). 
Our choice of these Coulomb integrals corresponds 
approximately to $U \sim 11$-12 eV, $U' \sim 10$ eV, 
and $J \sim 1$ eV, where $U$, $U'$ and $J$ are 
the intra-orbital, inter-orbital and Hund's couplings, respectively. 
This value of $U$ is similar to that in our previous study 
for copper oxides~\cite{Nomura2004,Nomura2005}. 
In addition, we take account of the Coulomb interaction 
between the $4p$ and $d$ electrons: 
$F_{pd}^0 = 3$ eV, $F_{pd}^2 = 3$ eV. 
For $H_{n.f.}$,  we determine the antiferromagnetic ground state 
within the HF approximation (see Appendix about details of HF calculation). 

\begin{figure}
\begin{center}
\includegraphics[width=90mm]{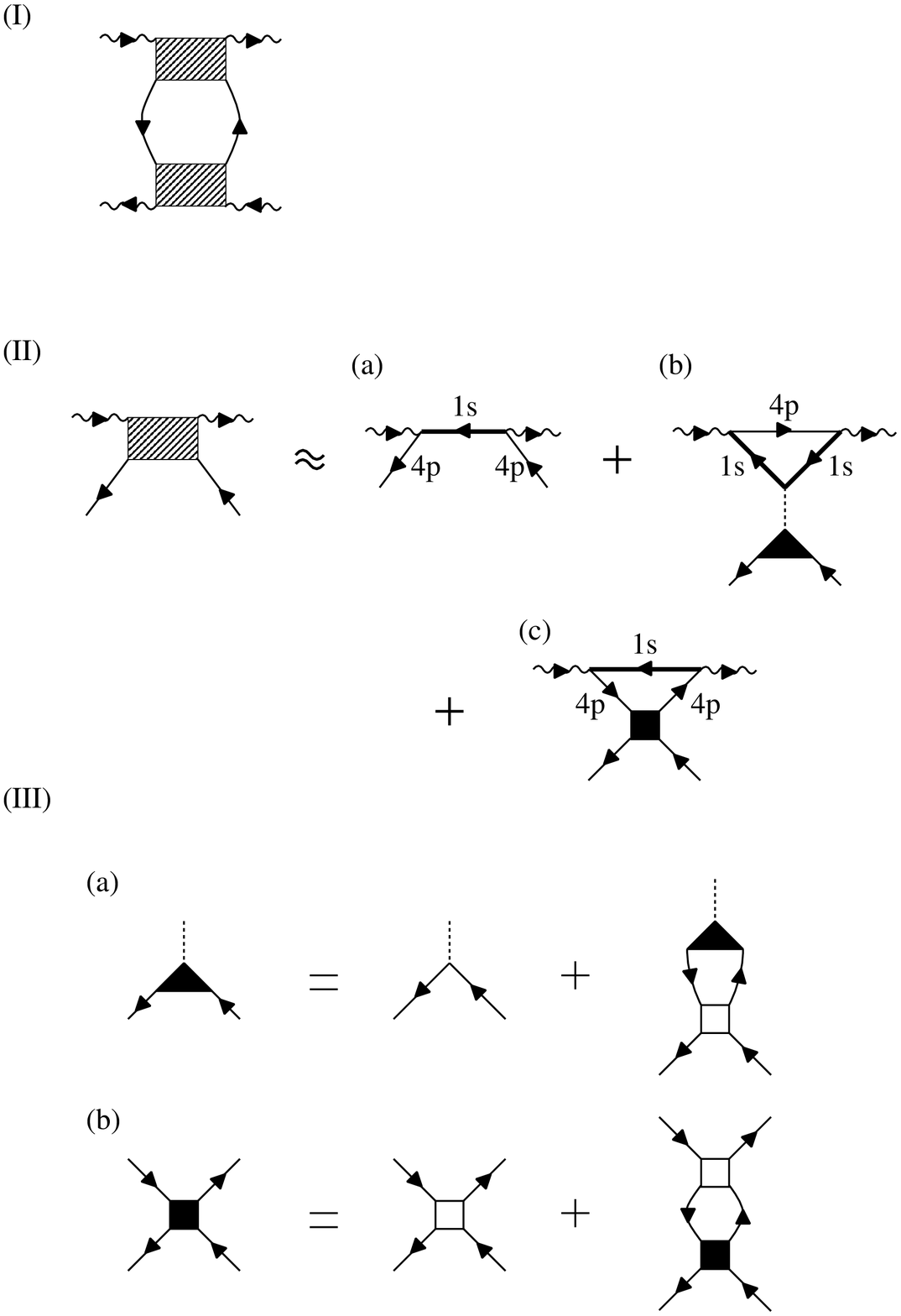}
\end{center}
\caption{
(I) RIXS intensity represented within the Keldysh perturbative formulation. 
The wavy lines and shaded rectangular represent 
the photon propagators and electron scattering vertex function $F(q, q')$, respectively. 
A pair of oriented solid lines represent the off-diagonal elements of the Keldysh Green's function, 
and connect the upper normally-time-ordered and lower reversely-time-ordered branches. 
(II) Approximate expansion for the scattering vertex function $F(q, q')$: 
(a) $F^{(0)}(q, q')$ for `0th-order process' (fluorescence), 
(b) $F^{(s)}(q, q')$ for `$s$-process', (c) $F^{(p)}(q, q')$ for `$p$-process'. 
The filled triangle and square are the three-point and four-point vertex functions 
to be renormalized by electron correlations, respectively. 
In (b), the dashed line represents the core-hole potential $V_{1s-d}$. 
Thick solid lines represent the propagator of the inner-shell $1s$ electrons. 
(III) RPA diagrams for the three-point and four-point vertex functions ((a) and (b), respectively), 
where empty squares represent the antisymmetrized bare Coulomb interaction 
$\Gamma^{(0)}$ among the $d$ and $4p$ electrons at transition-metal sites.} 
\label{Fig:diagrams}
\end{figure}

RIXS intensity can be obtained by calculating the number of photons generated 
in different states from the incident-photon state per unit time, 
as shown by Nozi\`eres and Abrahams~\cite{Nozieres1974}. 
To do this, we employ Keldysh perturbation theory 
as in Ref.~\ref{Nozieres1974} and our previous works~\cite{Nomura2004, Nomura2005, Igarashi2006}. 
The RIXS intensity is generally expressed by the diagram (I) in Fig.~\ref{Fig:diagrams}, 
if assuming that only a single electron-hole pair remains in the final state. 
The analytic expression of RIXS intensity is obtained from the diagram (I) 
of Fig.~\ref{Fig:diagrams} as: 
\begin{eqnarray}
W(q,q') &=& \frac{1}{N} \sum_{\mib{k}_1} \int_{-\infty}^{\infty} 
\frac{d \omega_1}{2\pi} \sum_{j_1j_2} G_{j_1}^+(k_1) G_{j_2}^-(k_1+Q) \nonumber \\
&& \times \biggl| \sum_{\mu\mu'}^{xyz} w_{\mu}(\mib{q}, \mib{e}) w_{\mu'}(\mib{q}', \mib{e}') 
F_{\mu\mu'; j_1j_2}(k_1; q, q') \biggr|^2, 
\label{Eq:W}
\end{eqnarray}
where $G_j^\pm(k)$ is the Keldysh Green's function~\cite{Keldysh1965}, 
$j_{1,2}$ are indices for the diagonalized bands, and $k_1 = (\omega_1, \mib{k}_1)$. 
$q$ and $q'$ are the four-momenta of the incident and emitted photons, respectively: 
$q = (\omega, \mib{q})$, $q' = (\omega', \mib{q}')$. $Q$ is the energy and momentum 
loss of the photon: $Q = q-q' = (\omega-\omega', \mib{q} - \mib{q}') \equiv (\Omega, \mib{Q})$. 
$F_{\mu\mu'; j_1j_2}(k_1; q, q')$ is the scattering vertex function expressed 
using only the usual causal electron Green's functions and electron-electron interaction. 
At this stage, we omit $\omega_1$ dependence of $F_{\mu\mu'; j_1j_2}(k_1; q, q')$, i.e., 
$F_{\mu\mu'; j_1j_2}(k_1; q, q') = F_{\mu\mu'; j_1j_2}(\mib{k}_1; q, q')$, 
because it is justified within the following approximation for $F_{\mu\mu'; j_1j_2}(k_1; q, q')$. 
Within the HF approximation, the Green's functions $G_j^\pm(k)$ are given by 
\begin{eqnarray}
G_j^+(k_1) &=& 2 \pi i n_j(\mib{k}_1) \delta(\omega_1-E_j(\mib{k}_1)), \label{Eq:G+}\\
G_j^-(k_1) &=& -2 \pi i [1-n_j(\mib{k}_1)] \delta(\omega_1-E_j(\mib{k}_1)), \label{Eq:G-}
\end{eqnarray}
where $E_j(\mib{k}_1)$ is the energy of diagonalized band $j$, 
and $n_j(\mib{k}_1)$ is the electron occupation density at momentum $\mib{k}_1$ in band $j$: 
$n_j(\mib{k}_1) = 1$ for $E_j(\mib{k}_1)<0$ and $n_j(\mib{k}_1) = 0$ for $E_j(\mib{k}_1)>0$. 
Substituting eqs.~(\ref{Eq:G+}) and (\ref{Eq:G-}) into eq.~(\ref{Eq:W}), we have 
\begin{eqnarray}
W(q,q') &=& \frac{2 \pi}{N} \sum_{\mib{k}_1} \sum_{j_1j_2} 
n_{j_1}(\mib{k}_1) [1-n_{j_2}(\mib{k}_1 + \mib{Q}) ]
\delta (\Omega + E_{j_1}(\mib{k}_1)- E_{j_2}(\mib{k}_1 + \mib{Q})) \nonumber \\
&& \times \biggl| \sum_{\mu\mu'}^{xyz} w_{\mu}(\mib{q}, \mib{e}) w_{\mu'}(\mib{q}', \mib{e}') 
F_{\mu\mu'; j_1j_2}(\mib{k}_1; q, q') \biggr|^2. 
\label{Eq:W2}
\end{eqnarray}
For calculation of $F_{\mu\mu'; j_1j_2}(\mib{k}_1; q, q')$, we use perturbation expansion 
with respect to electron-electron interactions. 
There are three major contributions to $F_{\mu\mu'; j_1j_2}(\mib{k}_1; q, q')$. 
The first is the zeroth-order term represented by the diagram (II)-(a) in Fig.~\ref{Fig:diagrams}. 
This diagram presents a main contribution to the fluorescence yield. 
We refer to this contribution as `0th-order process'. 
The second originates from the screening process of the $1s$ core hole. 
Within the Born approximation with respect to the core-hole potential $V_{1s-d}$, 
this process is expressed by the diagram (II)-(b) in Fig.~\ref{Fig:diagrams}. 
We refer to this contribution as `$s$-screening process' or `$s$-process'. 
The $s$-screening process has been included 
in our previous works~\cite{Nomura2004,Nomura2005}. 
The third describes the screening process of the excited $4p$ electron. 
Within the Born approximation (or equivalently the linear response approximation 
with respect to the potential polarizing the transition-metal $d$-electrons), 
this contribution is expressed by the diagram (II)-(c) in Fig.~\ref{Fig:diagrams}. 
We refer to this contribution as `$p$-screening process' or `$p$-process'. 
Of course, in higher-order contributions, more complex diagrams can appear, 
which cannot simply be classified to `$s$-screening process' or `$p$-screening process'. 
Nevertheless, this classification turns out to be convenient 
for microscopic analysis of RIXS spectra. 
Thus, we obtain the following approximate expression 
for the scattering vertex function: 
\begin{eqnarray}
F_{\mu\mu'; j_1j_2}(\mib{k}_1; q, q') &=& F^{(0)}_{\mu\mu'; j_1j_2}(\mib{k}_1; q, q') \nonumber\\
&& - \sum_{\zeta_1\zeta_2} u_{\zeta_2, j_2}^*(\mib{k}_1+\mib{Q}) u_{\zeta_1, j_1}(\mib{k}_1) 
[ F^{(s)}_{\mu\mu'; \zeta_1\zeta_2}(q, q') + F^{(p)}_{\mu\mu'; \zeta_1\zeta_2}(q, q') ], 
\label{Eq:F}
\end{eqnarray}
where $u_{\zeta, j}(\mib{k})$ is the diagonalization matrix 
of the HF Hamiltonian given by eq.~(\ref{Eq:HMF}). $\zeta_n$ is orbital-spin combined index: 
$\zeta_n = (\ell_n,\sigma_n)$, and $\sum_{\zeta_n} = \sum_{\ell_n}\sum_{\sigma_n} $, 
where $\ell_n$ represents $4p$ and $d$ orbitals at transition-metal sites. 
Contributions from the above three processes are given by 
\begin{eqnarray}
F^{(0)}_{\mu\mu'; j_1j_2}(\mib{k}_1; q, q') &=& \sum_i^{\rm t.m.u.} \sum_{\sigma} 
\frac{u_{4p_{\mu}(i)\sigma, j_2}^*(\mib{k}_1+\mib{Q})u_{4p_{\mu'}(i)\sigma, j_1}(\mib{k}_1)}
{\omega + \tilde{\varepsilon}_{1s}(\mib{r}_i) - E_{j_2}(\mib{k}_1+\mib{Q})}, \\
F^{(s)}_{\mu\mu'; \zeta_1\zeta_2}(q, q') &=& \sum_i^{\rm t.m.u.} V_{1s-d}(\mib{r}_i) 
\Lambda_{\zeta_2\zeta_1}(\mib{r}_i; Q) \nonumber\\
&&\times \sum_{j\sigma} \frac{1}{N} \sum_{\mib{k}}^{E>0} 
\frac{u_{4p_{\mu}(i)\sigma, j}^*(\mib{k}) u_{4p_{\mu'}(i)\sigma, j}(\mib{k})}
{[\omega + \tilde{\varepsilon}_{1s}(\mib{r}_i) - E_j(\mib{k})]
[\omega' + \tilde{\varepsilon}_{1s}(\mib{r}_i) - E_j(\mib{k})]}, \label{Eq:Fs} \\
F^{(p)}_{\mu\mu'; \zeta_1\zeta_2}(q, q') &=& \sum_i^{\rm t.m.u.} 
\sum_{\zeta_3\zeta_4}^{@\mib{r}_i} \Gamma_{\zeta_2\zeta_4,\zeta_3\zeta_1}(Q) \label{Eq:Fp}\\
&&\times \sum_{j_3j_4\sigma} \frac{1}{N} \sum_{\mib{k}}^{E>0} 
\frac{u_{\zeta_3, j_3}(\mib{k}+\mib{Q}) u_{4p_{\mu}(i)\sigma, j_3}^*(\mib{k}+\mib{Q}) 
u_{\zeta_4, j_4}^*(\mib{k}) u_{4p_{\mu'}(i)\sigma, j_4}(\mib{k})}
{[\omega + \tilde{\varepsilon}_{1s}(\mib{r}_i) - E_{j_3}(\mib{k}+\mib{Q})]
[\omega' + \tilde{\varepsilon}_{1s}(\mib{r}_i) - E_{j_4}(\mib{k})]}, \nonumber  
\end{eqnarray}
where $\Lambda_{\zeta_2\zeta_1}(\mib{r}_i; Q)$ and 
$\Gamma_{\zeta_2\zeta_4,\zeta_3\zeta_1}(Q)$ are the three-point and 
four-point vertex functions, which are represented by the filled triangle 
and square in Fig.~\ref{Fig:diagrams} (II) (b) and (c), respectively. 
$4p_\mu(i)\sigma$ means the $4p_\mu$ state at transition-metal 
site $\mib{r}_i$ with spin $\sigma$. 
$\tilde{\varepsilon}_{1s}(\mib{r}_i) \equiv \varepsilon_{1s}(\mib{r}_i) + i \Gamma_{1s} $, 
where $\Gamma_{1s}$ is the damping rate of the $1s$ core-hole 
and set to 0.8 eV in the present study. 
Summations in $i$ with `t.m.u.' at the top means that $\mib{r}_i$ 
should be restricted only to transition-metal sites in the unit cell. 
`$E>0$' appearing in the summation about $\mib{k}$ means restriction 
to the $\mib{k}$-region satisfying $E_j(\mib{k})>0$ in eq.~(\ref{Eq:Fs}), 
and to the $\mib{k}$-region satisfying both $E_{j_3}(\mib{k}+\mib{Q})>0$ 
and $E_{j_4}(\mib{k})>0$ in eq.~(\ref{Eq:Fp}). 
To obtain eq.~(\ref{Eq:Fp}), we have omitted the processes where, 
before the excited $4p$ electron interacts, 
the $1s$ core-hole annihilates with other $4p$ electrons. 
The omitted processes give only a behavior similar to usual fluorescence 
and is negligible in analysis of RIXS. 

The vertex functions introduced above are renormalized by electron correlations. 
We take account of electron correlations within RPA. 
RPA for $\Lambda_{\zeta_2\zeta_1}(\mib{r}_i; Q)$ and 
$\Gamma_{\zeta_2\zeta_4,\zeta_3\zeta_1}(Q)$ 
is represented diagrammatically in Fig.~\ref{Fig:diagrams} (III) (a) and (b), respectively. 
The analytic expressions for these diagrams are 
\begin{eqnarray}
\Lambda_{\zeta_2\zeta_1}(\mib{r}_i; Q) = \delta_{\zeta_1\zeta_2}^{{\rm t.m.}@\mib{r}_i}
- \sum_{\zeta_1'\zeta_2'}^{@\mib{r}_i} \sum_{\zeta_3'\zeta_4'} 
\Lambda_{\zeta_2'\zeta_1'}(\mib{r}_i; Q) \chi_{\zeta_3'\zeta_2',\zeta_1'\zeta_4'}(Q) 
\Gamma_{\zeta_2\zeta_4';\zeta_3'\zeta_1}^{(0)}, \label{Eq:Lambda}\\ 
\Gamma_{\zeta_2\zeta_4;\zeta_3\zeta_1}(Q) = \Gamma_{\zeta_2\zeta_4;\zeta_3\zeta_1}^{(0)} 
- \sum_{\zeta_1'\zeta_2'} \sum_{\zeta_3'\zeta_4'} 
\Gamma_{\zeta_2'\zeta_4;\zeta_3\zeta_1'}(Q) \chi_{\zeta_3'\zeta_2',\zeta_1'\zeta_4'}(Q) 
\Gamma_{\zeta_2\zeta_4';\zeta_3'\zeta_1}^{(0)}, \label{Eq:Gamma}
\end{eqnarray}
where $\Gamma_{\zeta_1\zeta_2;\zeta_3\zeta_4}^{(0)}$ is 
the antisymmetrized bare Coulomb interaction given by 
$\Gamma_{\zeta_1\zeta_2;\zeta_3\zeta_4}^{(0)} = 
I_{\ell_1\ell_2;\ell_3\ell_4}\delta_{\sigma_1\sigma_4}\delta_{\sigma_2\sigma_3} 
- I_{\ell_1\ell_2;\ell_4\ell_3}\delta_{\sigma_1\sigma_3}\delta_{\sigma_2\sigma_4}$, 
and $\delta_{\zeta_1\zeta_2}^{{\rm t.m.}@\mib{r}_i} = \delta_{\zeta_1\zeta_2} = 
\delta_{\ell_1\ell_2} \delta_{\sigma_1 \sigma_2}$ 
only when both of the orbitals $\ell_1$ and $\ell_2$ are placed on the transition-metal 
site $\mib{r}_i$, and otherwise  $\delta_{\zeta_1\zeta_2}^{{\rm t.m.}@\mib{r}_i} = 0$. 
$\chi(Q)$ is the polarization function calculated by 
\begin{eqnarray}
\chi_{\zeta_3\zeta_2,\zeta_1\zeta_4}(Q) &=& \frac{1}{N} \sum_{\mib{k}} \sum_{jj'} 
u_{\zeta_1, j}(\mib{k}) u_{\zeta_4, j}^*(\mib{k})
u_{\zeta_3, j'}(\mib{k}+\mib{Q}) u_{\zeta_2, j'}^*(\mib{k}+\mib{Q}) \chi_{jj'}(\mib{k}; Q), \\ 
\chi_{jj'}(\mib{k}; Q) &=& \frac{n_{j'}(\mib{k}+\mib{Q}) - n_j(\mib{k})}
{\Omega + E_j(\mib{k}) - E_{j'}(\mib{k} + \mib{Q}) + i \Gamma_{eh}}, 
\end{eqnarray}
where $\Gamma_{eh}$ is interpreted as the damping rate 
of the excited electron-hole pair near the Fermi level. 
Solving eqs.~(\ref{Eq:Lambda}) and (\ref{Eq:Gamma}), we can determine 
$\Lambda_{\zeta_2\zeta_1}(\mib{r}_i; Q)$ 
and $\Gamma_{\zeta_2\zeta_4,\zeta_3\zeta_1}(Q)$ within RPA. 

To resolve contributions from each process of the 0th-order, $s$-screening 
and $p$-screening, we introduce the process-resolved spectra as follows: 
\begin{eqnarray}
W^{(0)}(q,q') &=& \frac{2 \pi}{N} \sum_{\mib{k}_1} \sum_{j_1j_2} 
n_{j_1}(\mib{k}_1) [1-n_{j_2}(\mib{k}_1 + \mib{Q}) ]
\delta (\Omega + E_{j_1}(\mib{k}_1)- E_{j_2}(\mib{k}_1 + \mib{Q})) \nonumber \\
&& \times \biggl| \sum_{\mu\mu'}^{xyz} w_{\mu}(\mib{q}, \mib{e}) w_{\mu'}(\mib{q}', \mib{e}') 
F_{\mu\mu'; j_1j_2}^{(0)}(\mib{k}_1; q, q') \biggr|^2, \\ 
W^{(s)}(q,q') &=& \frac{2 \pi}{N} \sum_{\mib{k}_1} \sum_{j_1j_2} 
n_{j_1}(\mib{k}_1) [1-n_{j_2}(\mib{k}_1 + \mib{Q}) ]
\delta (\Omega + E_{j_1}(\mib{k}_1)- E_{j_2}(\mib{k}_1 + \mib{Q})) \nonumber \\
&& \times \biggl| \sum_{\mu\mu'}^{xyz} w_{\mu}(\mib{q}, \mib{e}) w_{\mu'}(\mib{q}', \mib{e}') 
\sum_{\zeta_1\zeta_2} u_{\zeta_2, j_2}^*(\mib{k}_1+\mib{Q})u_{\zeta_1, j_1}(\mib{k}_1) 
F_{\mu\mu'; \zeta_1\zeta_2}^{(s)}(q, q') \biggr|^2, \label{Eq:Ws}\\ 
W^{(p)}(q,q') &=& \frac{2 \pi}{N} \sum_{\mib{k}_1} \sum_{j_1j_2} 
n_{j_1}(\mib{k}_1) [1-n_{j_2}(\mib{k}_1 + \mib{Q}) ]
\delta (\Omega + E_{j_1}(\mib{k}_1)- E_{j_2}(\mib{k}_1 + \mib{Q})) \nonumber \\
&& \times \biggl| \sum_{\mu\mu'}^{xyz} w_{\mu}(\mib{q}, \mib{e}) w_{\mu'}(\mib{q}', \mib{e}') 
\sum_{\zeta_1\zeta_2} u_{\zeta_2, j_2}^*(\mib{k}_1+\mib{Q})u_{\zeta_1, j_1}(\mib{k}_1) 
F_{\mu\mu'; \zeta_1\zeta_2}^{(p)}(q, q') \biggr|^2. \label{Eq:Wp}
\end{eqnarray}
These are obtained from eq.~(\ref{Eq:W2}) by keeping only one of 
$F^{(0)}_{\mu\mu'; j_1j_2}(\mib{k}_1; q, q')$, $F^{(s)}_{\mu\mu'; \zeta_1\zeta_2}(q, q')$ and 
$F^{(p)}_{\mu\mu'; \zeta_1\zeta_2}(q, q')$ and setting the rest two to zero in eq.~(\ref{Eq:F}). 

Further to resolve orbital-excitation processes involved in the $s$-screening 
and $p$-screening processes, we introduce the orbital-resolved spectra as follows: 
\begin{eqnarray}
W_{\ell_1 \rightarrow \ell_2}^{(s)}(q,q') &=& \frac{2 \pi}{N} \sum_{\mib{k}_1} \sum_{j_1j_2} 
n_{j_1}(\mib{k}_1) [1-n_{j_2}(\mib{k}_1 + \mib{Q}) ]
\delta (\Omega + E_{j_1}(\mib{k}_1)- E_{j_2}(\mib{k}_1 + \mib{Q})) \nonumber \\
&& \times \biggl| \sum_{\mu\mu'}^{xyz} w_{\mu}(\mib{q}, \mib{e}) w_{\mu'}(\mib{q}', \mib{e}') 
\sum_{\sigma_1\sigma_2} u_{\zeta_2, j_2}^*(\mib{k}_1+\mib{Q})u_{\zeta_1, j_1}(\mib{k}_1) 
F_{\mu\mu'; \zeta_1\zeta_2}^{(s)}(q, q') \biggr|^2, \label{Eq:Wsll}\\ 
W_{\ell_1 \rightarrow \ell_2}^{(p)}(q,q') &=& \frac{2 \pi}{N} \sum_{\mib{k}_1} \sum_{j_1j_2} 
n_{j_1}(\mib{k}_1) [1-n_{j_2}(\mib{k}_1 + \mib{Q}) ]
\delta (\Omega + E_{j_1}(\mib{k}_1)- E_{j_2}(\mib{k}_1 + \mib{Q})) \nonumber \\
&& \times \biggl| \sum_{\mu\mu'}^{xyz} w_{\mu}(\mib{q}, \mib{e}) w_{\mu'}(\mib{q}', \mib{e}') 
\sum_{\sigma_1\sigma_2} u_{\zeta_2, j_2}^*(\mib{k}_1+\mib{Q})u_{\zeta_1, j_1}(\mib{k}_1) 
F_{\mu\mu'; \zeta_1\zeta_2}^{(p)}(q, q') \biggr|^2, \label{Eq:Wpll} 
\end{eqnarray}
where the orbital indices $\ell_1$ and $\ell_2$ specify the initial and final orbitals 
excited in RIXS, respectively. 
Equations (\ref{Eq:Wsll}) and (\ref{Eq:Wpll}) are obtained by suspending the summation 
with respect to orbital indices $\ell_1$ and $\ell_2$ in the right-hand side 
of eqs.~(\ref{Eq:Ws}) and ~(\ref{Eq:Wp}). 

Here we should note that the total RIXS intensity $W(q, q')$ does not equal the sum 
of the resolved intensities, e.g., $W(q, q') \neq W^{(0)}(q, q') + W^{(s)}(q, q') + W^{(p)}(q, q')$, 
$W^{(s)}(q, q') \neq \sum_{\ell\ell'} W_{\ell \rightarrow \ell'}^{(s)}(q, q')$, 
$W^{(p)}(q, q') \neq \sum_{\ell\ell'} W_{\ell \rightarrow \ell'}^{(p)}(q, q')$, and so on. 
This is because the total summed spectrum $W(q, q')$ contains interference terms 
such as $F^{(s)}(q, q') F^{(p)}{}^*(q, q')$, while $W^{(s)}(q, q')$ and $W^{(p)}(q, q')$ 
contain only $|F^{(s)}(q, q')|^2$ and $|F^{(p)}(q, q')|^2$, respectively. 
Nevertheless, the resolved spectra introduced above turn out to be convenient 
for microscopic analysis of RIXS spectra, as we see in the next section. 

For numerical calculation of eq.~(\ref{Eq:W2}), 
we use the Lorentzian expression for the $\delta$-function: 
\begin{equation}
\delta (\Omega + E_{j_1}(\mib{k}_1)- E_{j_2}(\mib{k}_1 + \mib{Q})) \rightarrow 
\frac{1}{\pi} \frac{\epsilon}
{[\Omega + E_{j_1}(\mib{k}_1)- E_{j_2}(\mib{k}_1 + \mib{Q})]^2 + \epsilon^2},  
\end{equation}
where $\epsilon$ is usually a small positive factor. This function possesses poles 
at $\Omega = E_{j_2}(\mib{k}_1 + \mib{Q}) - E_{j_1}(\mib{k}_1) \pm i \epsilon$, 
which correspond to the transition from band $j_1$ to band $j_2$. 
Therefore, at a first glance, one might consider that eq.~(\ref{Eq:W2}) 
describes only simple band-to-band transitions and fails to describe local $d$-$d$ transitions. 
This naive view is not correct, as explained next. 
We should note that, for overall consistency, the factor $\epsilon$ should 
equals $\Gamma_{eh}$, where $\Gamma_{eh}$ is the damping rate 
of excited electron-hole pair, already introduced above. 
Setting $\epsilon \equiv \Gamma_{eh}$, the position of the pole 
$\Omega = E_{j_2}(\mib{k}_1 + \mib{Q}) - E_{j_1}(\mib{k}_1) \pm i \Gamma_{eh}$ 
is modified to a non-trivial position by the RPA correction. 
The modified poles describe bound states 
between the excited electron and hole in the final state. 
In fact, as we see in the next section, not only charge-transfer excitations 
but also local $d$-$d$ excitations can be described within our HF-RPA calculation. 
In the present study, we take $\epsilon \equiv \Gamma_{eh} = 20$ meV. 
Here we should note that sharpness of RIXS spectra is determined 
by $\Gamma_{eh}$, not by $\Gamma_{1s}$. 

\section{Numerical Results}
\label{Sc:Results}

\begin{figure}
\begin{center}
\includegraphics[width=90mm]{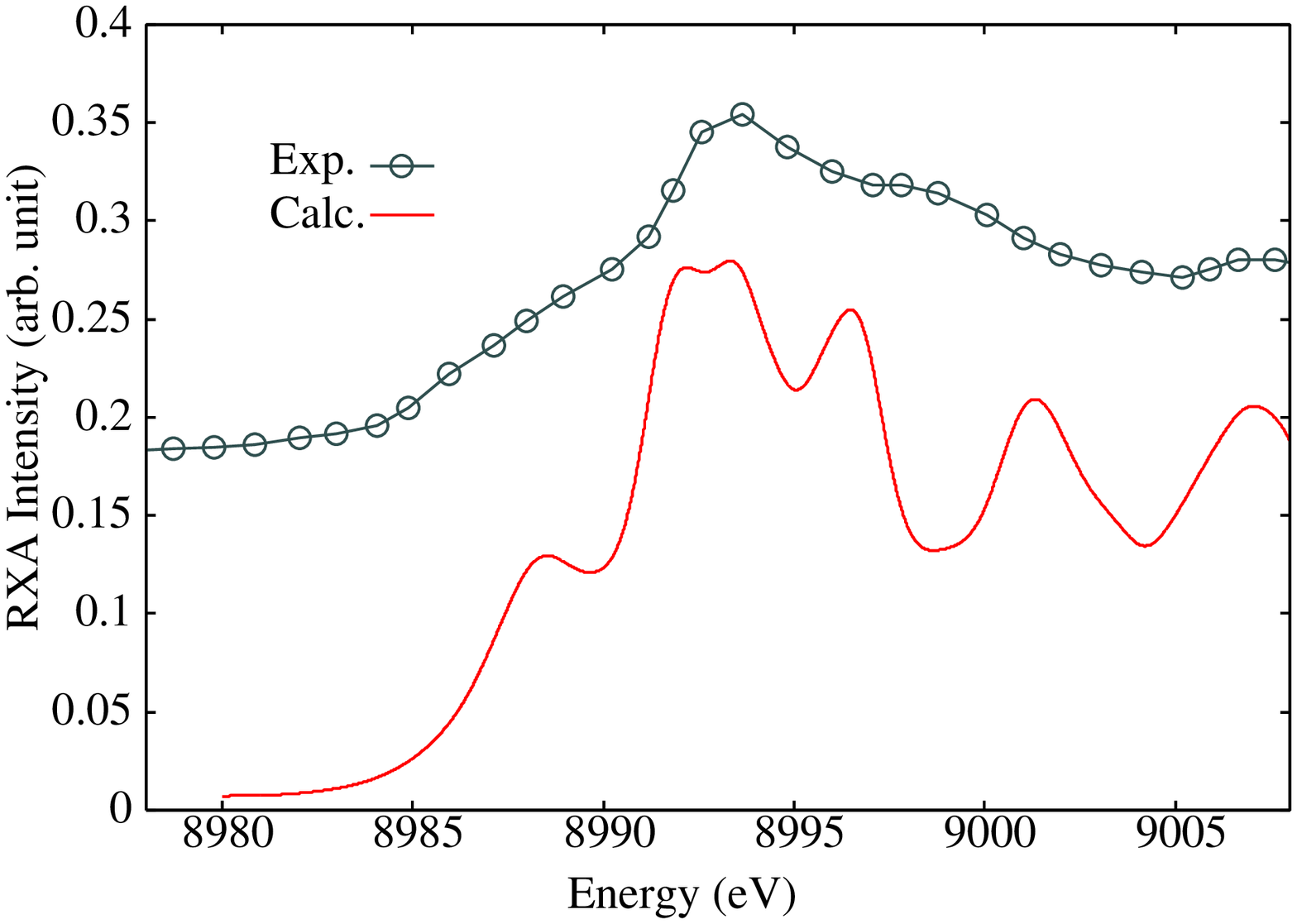}
\end{center}
\caption{(Color online)
Circles connected by a line represent the experimental RXA spectrum 
read from Ref.~\ref{Caciuffo2002}, and the solid curve represents the calculated result.}
\label{Fig:rxas}
\end{figure}

In order to set the Cu-$1s$ energy level $\varepsilon_{1s}$, 
we calculate the resonant x-ray absorption (RXA) spectra using 
\begin{eqnarray}
I_{\rm RXA}(\omega) &=& 2 \pi \sum_{\mu}^{xyz} |w_\mu({\mib q},{\mib e}) |^2 
\rho_{4p_\mu} (\omega + \varepsilon_{1s})  \nonumber\\
& \sim & \sum_{\mu}^{xyz} \rho_{4p_\mu} (\omega + \varepsilon_{1s}) \nonumber\\
& \propto &  \rho_{4p} (\omega + \varepsilon_{1s}) 
\end{eqnarray}
where $\rho_{4p_\mu} (\omega) $ and $\rho_{4p} (\omega) $ 
are the partial density of states of the $4p_{\mu}$ orbital 
and the total density of states of the $4p$ orbitals, respectively. 
Here we have neglected the influence of the core hole, 
and the $4p$ density of states $\rho_{4p} (\omega)$ is calculated 
within the band structure calculation. 
In Fig.~\ref{Fig:rxas}, calculated and experimental RXA spectra are compared 
with each other (The Lorentzian broadening factor is set to $\Gamma_{1s} = 0.8$ eV). 
From consistency about the main peak position, we set $\varepsilon_{1s} = -8980$ eV. 

\begin{figure}
\begin{center}
\includegraphics[width=70mm]{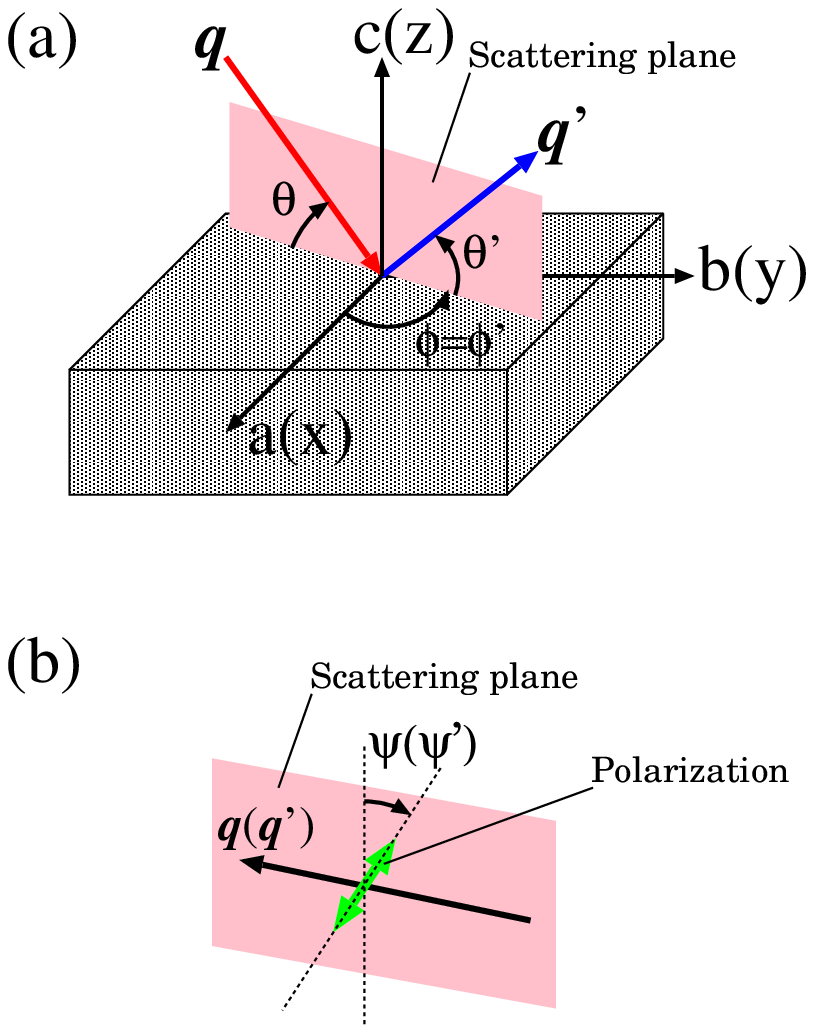}
\end{center}
\caption{
(Color online)
Definitions of angles characterizing the scattering geometry. 
$\mib{q}$ and $\mib{q}$' are the momentum vectors 
of the incident and emitted photons. 
$\theta$,  $\phi$ and $\psi$ ($\theta$',  $\phi$' and $\psi$') 
are the Bragg, azimuthal and polarization angles, respectively, 
for incident (emitted) photons. 
The $a$- and $c$-axes are parallel along those of the crystalline lattice. 
The polarization angle is measured with respect to the scattering plane, 
i.e., $\psi=0$ ($\psi=\pi/2$) means that the polarization direction 
is parallel (perpendicular) to the scattering plane.}
\label{Fig:angles}
\end{figure}

Here we define the angles characterizing the scattering geometry as shown in Fig.~\ref{Fig:angles}. 
We take the following parameters for numerical calculations: 
$\psi =0$ rad (i.e., incident photons are in $\pi$-polarization), 
$\theta = \theta' = 0.24 \pi$ rad, $\phi = \phi' = 0$ rad. 
The incident photon energy is fixed to $\omega = 8994$ eV, 
as in the experiment~\cite{Ishii2011}. 

\begin{figure}
\begin{center}
\includegraphics[width=90mm]{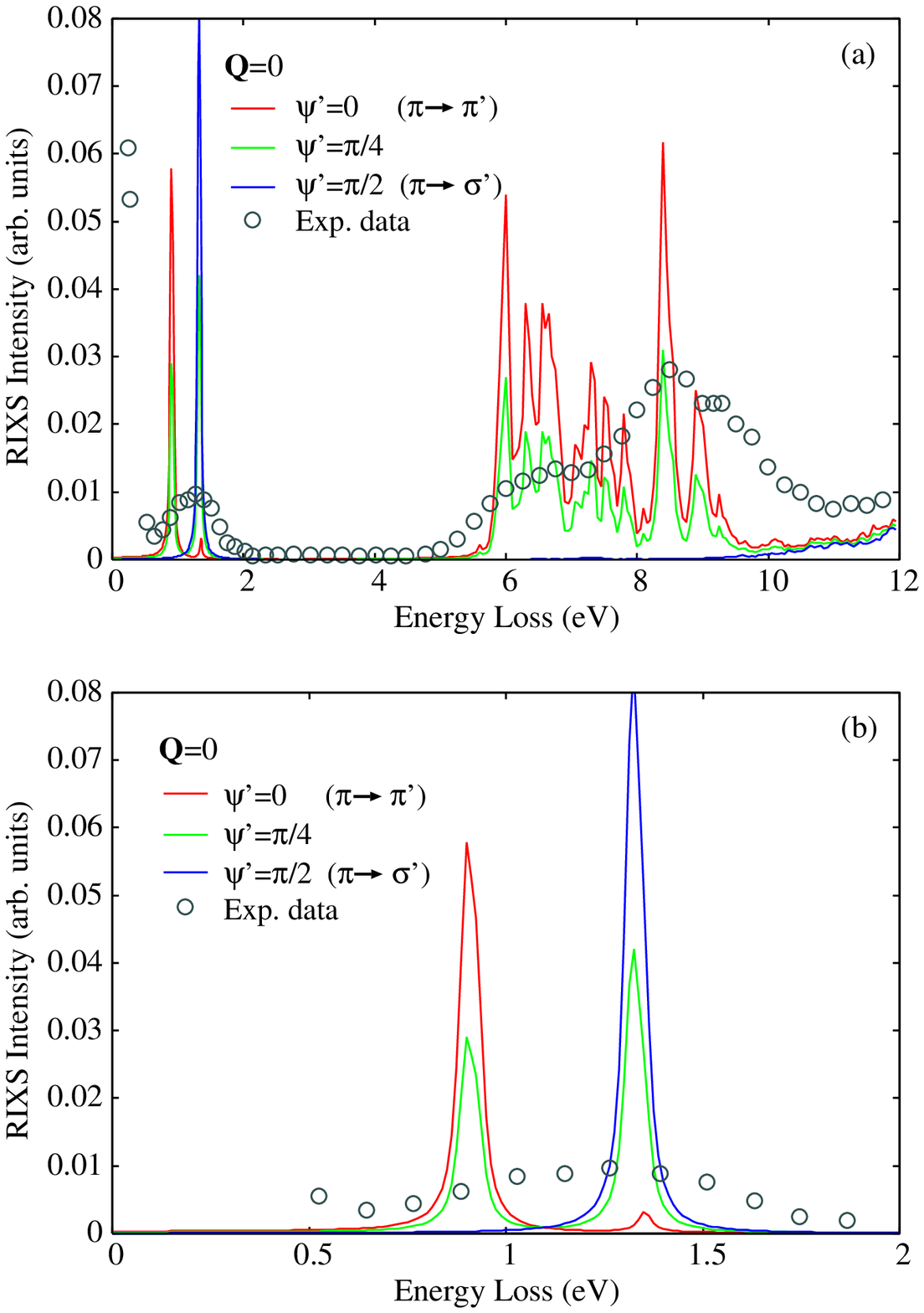}
\end{center}
\caption{
(Color online) 
Polarization dependence of calculated RIXS spectra 
and comparison with typical experimental data. 
Solid circles are the experimental data read 
from Ref.~\ref{Ishii2011} (not polarization-resolved). 
Momentum transfer of the photon is set to the $\Gamma$ point: 
$\mib{Q} = \mib{q}-\mib{q}' = 0$. 
In (b), the low-energy region is enlarged.  }
\label{Fig:poldeptot}
\end{figure}

Typical calculated results of the total RIXS intensity $W(q, q')$ 
are compared with typical experimental data in Fig.~\ref{Fig:poldeptot}. 
Roughly speaking, there are two characteristic features: 
low-energy feature around 1-2 eV and high-energy feature above 5 eV. 
It becomes clear below that the low-energy feature originates 
from the $d$-$d$ excitations among the Cu-$d$ orbitals, 
supporting the interpretation in Ref.~\ref{Ishii2011}. 
On the other hand, the high-energy feature is mainly attributed 
to the charge-transfer excitations between Cu-$d$ and F-$p$ states, 
as understood from the electronic structure in Fig.~\ref{Fig:bandsdos}. 
Both of the features show notable polarization dependence. 
Particularly, the ratio of the peak intensity around 0.9 eV and 1.4 eV 
is drastically changed, as the polarization direction of emitted photons 
is changed from $\pi$' ($\psi' = 0$) to $\sigma$' ($\psi' = \pi /2$), 
as seen in Fig.~\ref{Fig:poldeptot} (b). 
This behavior is qualitatively consistent 
with experimental data in Ref.~\ref{Ishii2011}. 

\begin{figure}
\begin{center}
\includegraphics[width=140mm]{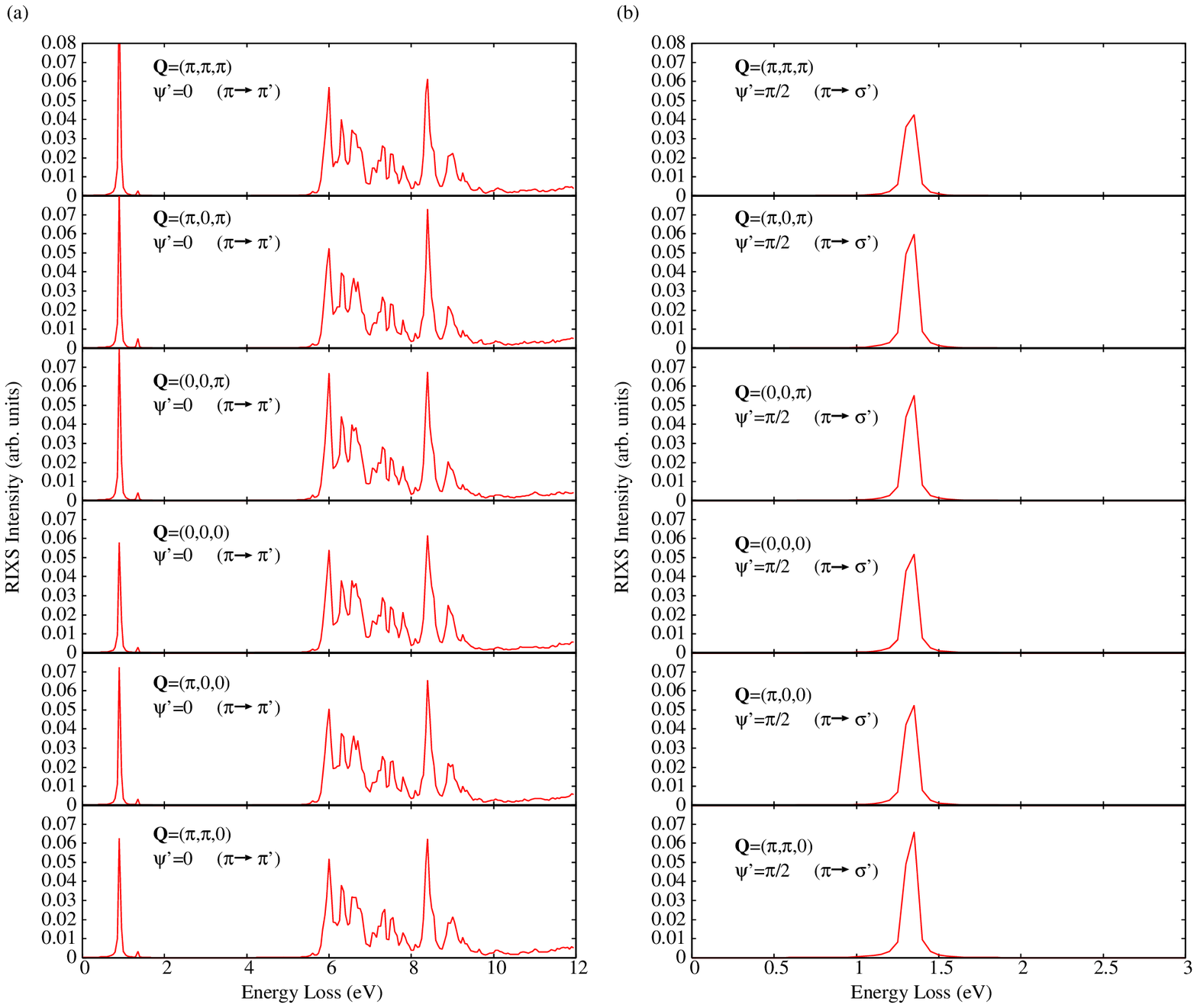}
\end{center}
\caption{
(Color online)
Momentum dependence of the calculated RIXS spectra 
for two cases of polarization: 
(a) $\pi \rightarrow \pi'$ ($\psi = 0$, $\psi' = 0$), 
(b) $\pi \rightarrow \sigma'$ ($\psi = 0$, $\psi' = \pi/2$). 
In (b), only the low-energy weights are shown, 
because the high-energy weights are almost 
suppressed in that scattering geometry. }
\label{Fig:mdeptot}
\end{figure}

Momentum dependence of calculated RIXS spectra is shown in Fig.~\ref{Fig:mdeptot}. 
For both the cases of $\pi'$- and $\sigma'$- polarizations, 
RIXS spectra do not exhibit notable momentum dependence 
all over the region of energy loss. 
This suggests that the excitations related to these RIXS weights are spatially localized. 
This is in strong contrast to the cases of copper oxides, 
where RIXS weights show strong characteristic 
momentum dependence~\cite{Hasan2000,Kim2002,Ishii2005}. 
The absence of notable momentum dependence in KCuF$_3$ 
is reasonably understood, since the relevant bands are rather flat, 
i.e., do not strongly depend on momentum, as shown 
in Fig.~\ref{Fig:bandsdos}(a). 

\begin{figure}
\begin{center}
\includegraphics[width=140mm]{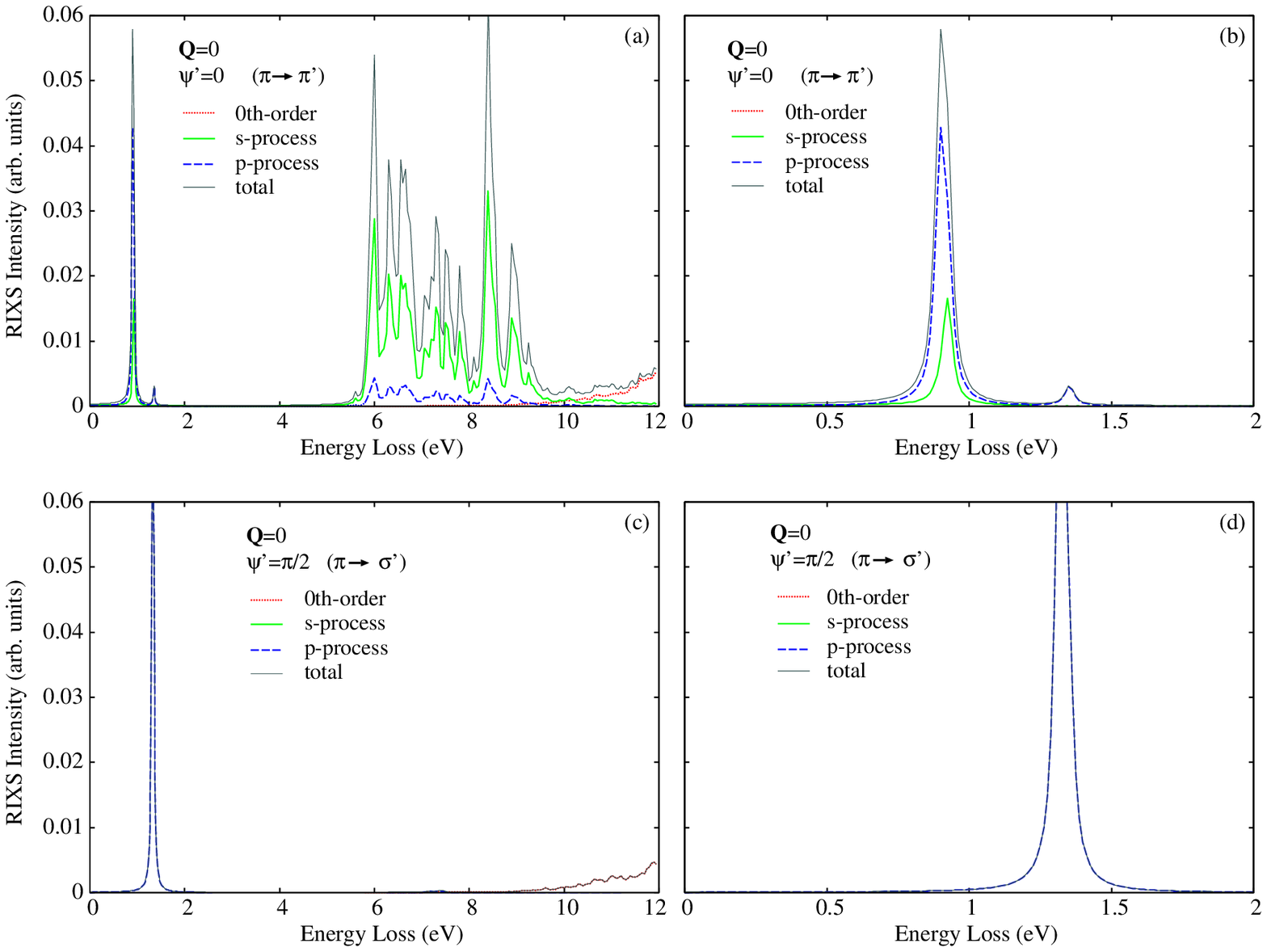}
\end{center}
\caption{
(Color online)
Process-resolved RIXS weights for two cases of polarization. 
The low-energy region of each left-hand panel is enlarged 
in the corresponding right-hand panel. 
Momentum transfer of the photon is set to $\mib{Q} = \mib{q}-\mib{q}' = 0$. 
In (c), the 0th-order and total spectra give an almost identical curve above 8 eV. 
In (c) and (d), the $p$-process and total spectra give an almost identical curve 
below 2 eV. 
}
\label{Fig:prcrestot}
\end{figure}

To elucidate the microscopic origin of each RIXS weight, 
we present numerical results for the process-resolved spectra defined in the last section. 
Calculated results of $W^{(0)}(q,q')$ (0th-order),  $W^{(s)}(q,q')$ ($s$-process) 
and $W^{(p)}(q,q')$ ($p$-process) are presented in Fig.~\ref{Fig:prcrestot}. 
From Fig.~\ref{Fig:prcrestot}(a) and (b), we can see that 
the low-energy features around 0.9 eV and 1.4 eV are attributed 
to the $s$-process and $p$-process, 
while the high-energy charge-transfer weight originates mainly from the $s$-process. 
The tendency to increase above 10 eV is attributed to the 0th-order $W^{(0)}(q, q')$, 
and therefore is considered as the tail of the fluorescence yield. 
Concerning the low-energy features, the peak feature around 0.9 eV 
is induced through both the $s$-process and $p$-process,  
while the peak feature around 1.4 eV is induced only through the $p$-process, 
as seen from Fig. ~\ref{Fig:prcrestot}. 

\begin{figure}
\begin{center}
\includegraphics[width=140mm]{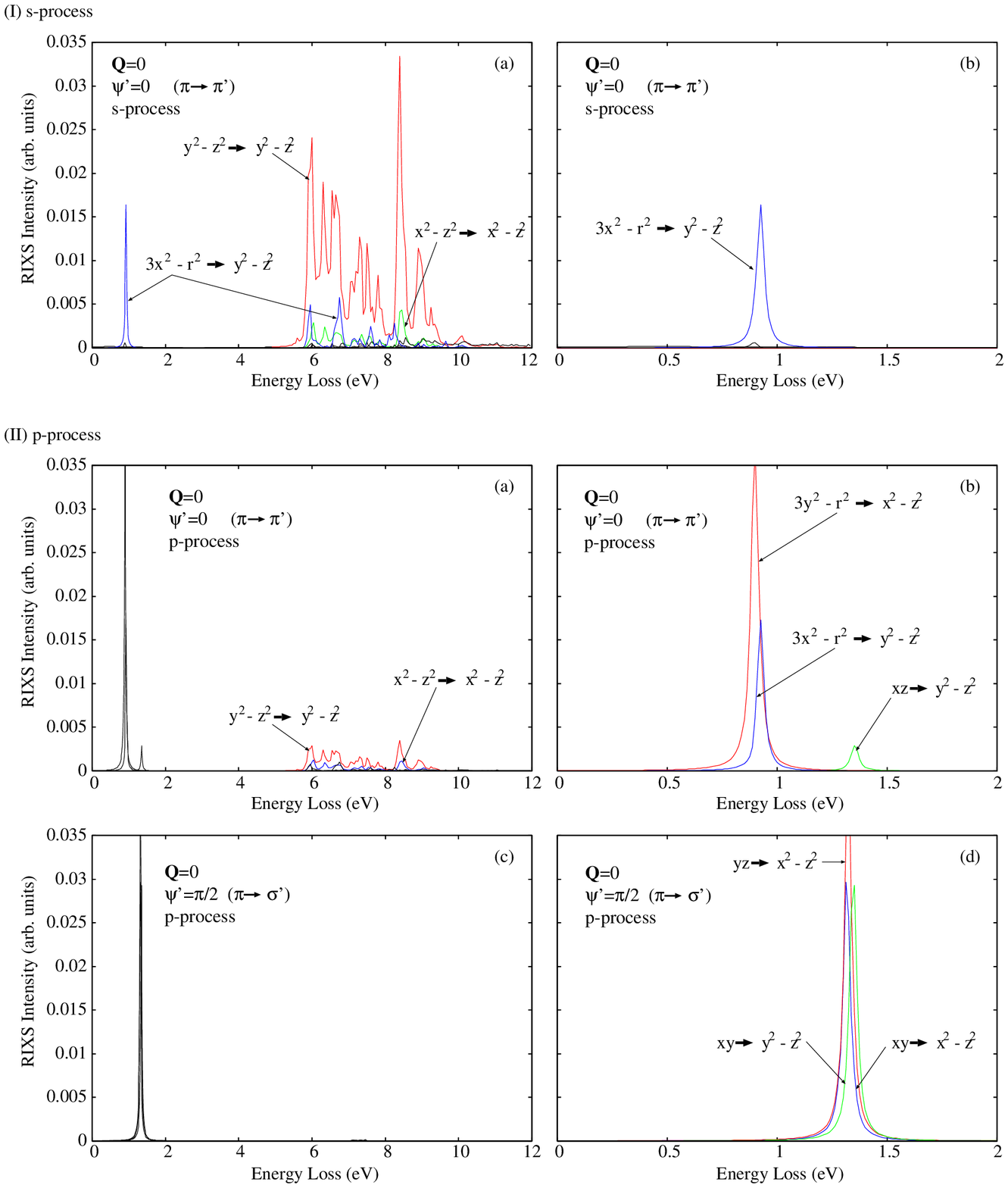}
\end{center}
\caption{
(Color online)
Orbital-resolved RIXS weights. 
(I)-(a) and (b) Orbital-resolved weights in the $s$-process, $W_{\ell \rightarrow \ell'}^{(s)}(q, q')$. 
(II)-(a)-(d) Orbital resolved weights in the $p$-process, $W_{\ell \rightarrow \ell'}^{(p)}(q, q')$. 
The low-energy region of each left-hand panel is enlarged in the corresponding right-hand panel. 
In the $s$-process (I), the results for $\pi \rightarrow \sigma'$ 
are not displayed, because the weights are almost suppressed to a negligible magnitude. 
Momentum transfer of the photon is set to $\mib{Q} = \mib{q}-\mib{q}' = 0$. }
\label{Fig:wrestot}
\end{figure}

To inspect RIXS weights more microscopically, 
we proceed to the calculated results of orbital-resolved spectra. 
Calculated orbital-resolved RIXS spectra $W_{\ell \rightarrow \ell'}^{(s,p)}(q, q')$
are presented in Fig.~\ref{Fig:wrestot}, 
where all the contributions from possible 128 orbital-excitation processes 
(from 8 orbitals to 8 orbitals at each of two Cu sites in the unit cell) are plotted. 
In the $s$-process, relevant excitations occur only among the $d\gamma$ ($e_g$) orbitals, 
as seen in Fig.~\ref{Fig:wrestot} (I). 
It is remarkable that the aspect of orbital excitations is very different 
between the low-energy and high-energy regions: off-diagonal orbital excitations 
($W_{\ell \rightarrow \ell'}^{(s)}$ with $\ell \neq \ell'$) are dominant 
in the low-energy region, while only diagonal orbital excitations 
($W_{\ell \rightarrow \ell}^{(s)}$) are dominant in the high-energy region. 
This property also holds for the $p$-process, as seen in Fig.~\ref{Fig:wrestot} (II). 
Focusing on the low-energy region (see the right-hand side of Fig.~\ref{Fig:wrestot}), 
we see that the RIXS weight around 0.9 eV is attributed to the orbital excitations 
among the $d\gamma$ ($e_g$) orbitals, and that around 1.4 eV is attributed 
to the orbital excitations from the $d\varepsilon$ ($t_{2g}$) orbitals 
to the $d\gamma$ ($e_g$) orbitals. 
This result is consistent with the previous simple phenomenological 
assignment based on symmetry properties~\cite{Ishii2011}. 

\section{Discussions}
\label{Sc:Discussions}

In this section, we present some remarks on the formulation and calculated results. 

To explain the polarization dependence of RIXS spectra, 
we have included the $p$-process as well as the $s$-process. 
It should be noted that without the $p$-process, 
we could not explain the experimental spectra 
at the scattering geometry $\pi \rightarrow \sigma'$: 
the contributions through the $s$-process to the low-energy weights around 1-2 eV 
are almost completely suppressed at $\pi \rightarrow \sigma'$, 
as seen in Fig~\ref{Fig:prcrestot}(d), which is inconsistent with the experiment. 
Thus it is suggested that the $p$-process essentially occurs in RIXS of KCuF$_3$. 
In the $p$-process, the Coulomb interaction $F_{pd}$ between the $p$ and $d$ orbitals 
at Cu sites plays an essential role. 
To our knowledge, the effect of $F_{pd}$ on the polarization dependence 
was discussed theoretically for the first time by Ishihara 
in the case of copper oxides~\cite{Ishihara2008}. 
It may be considered that our present work is a practical application 
of their mechanism to a more realistic and complex electronic structure. 

To study orbital-excitation processes microscopically in detail, 
we have introduced orbital-resolved spectra. 
Such analysis has already been applied to RIXS at the Fe $K$ edge 
in iron-pnictide superconductors~\cite{Jarrige2012}.  
In iron pnictides, diagonal orbital excitations are dominant, 
and off-diagonal ones are almost irrelevant. 
In this sense, the orbital excitations in KCuF$_3$ 
are substantially different from those in iron pnictides. 
Roughly speaking, this may be because the symmetry 
of atom configuration around transition-metal sites 
is lower in KCuF$_3$ than in iron pnictides. 

Our orbital-resolving analysis suggests that the low-energy features around 1-2 eV 
originate from the $d$-$d$ excitations among the Cu-$d$ orbitals. 
These weights do not show any notable momentum dependence 
as shown in Fig.~\ref{Fig:mdeptot} and Ref.~\ref{Ishii2011}. 
Therefore, we should not regard them as a manifestation of orbital waves 
(or the so-called `orbitons'). 
Orbital waves should show some dispersive behavior as usual collective modes, 
if they were indeed observed. 

One might consider that the calculated spectra are much sharp 
and show fine structures, which were not observed experimentally. 
Sharpness of the calculated spectra depends on the electron-hole 
damping rate ($\Gamma_{eh}=20$ meV in the present work). 
If we take a larger damping rate, then those sharp and fine structures 
could be smeared to be broad peaks and possibly become similar 
to the experimental data. 
However, such fine structures as obtained in our present calculation 
could become observable in future experiments if the resolution is improved. 

In the HF calculation, we have modified one-particle energy levels 
by subtraction (see Appendix). 
Without this subtraction, we can still obtain almost the same charge-transfer weights 
above 5 eV, but no longer obtain the low-energy weights around 1-2 eV. 
They disappear to the negative side on the energy-loss axis. 
At present, we consider that the LDA band structure calculation, 
the tight-binding fitting or the HF approximation may not be sufficiently 
precise to evaluate the one-particle energy levels, because the one-particle energy 
levels are possibly much more influenced by on-site electron correlations 
than the hoppings between different sites are. 
Thus, we consider that precise evaluation of the one-particle energy levels 
is still difficult, while the hoppings are precisely evaluated. 

\section{Conclusions}
\label{Sc:Conclusions}

We have microscopically discussed RIXS at the Cu $K$ edge 
in a typical orbital-ordered compound KCuF$_3$. 
In our previous works~\cite{Nomura2004,Nomura2005}, 
we have taken account of only the `$s$-process', 
where the $1s$ core hole created in the intermediate state 
is screened by the Cu-$d$ electrons. 
However, the previous theoretical framework is insufficient 
to explain the experimental results, particularly, 
the polarization dependence in KCuF$_3$. 
We have shown that to explain the polarization dependence, 
the `$p$-process' plays an essential role, where the $4p$ electron excited 
in the intermediate state is screened by the Cu-$d$ electrons, 
in other words, the $4p$ electron scatters the Cu-$d$ electrons in the $p$-channel. 

To analyze further the RIXS process microscopically, 
we have introduced a new method of orbital-resolving analysis. 
This method enables us to clarify which orbital excitation 
is responsible for each spectral weight. 
As a result of our microscopic orbital-resolving analysis, 
high-energy spectral weights (above 5 eV) originate from charge-transfer excitations 
related to the Cu-$d\gamma$ orbitals, while the low-energy weights (below 2 eV) 
originate from the $d$-$d$ orbital excitations among the five Cu-$d$ orbitals. 
Thus we have succeeded in assigning specifically the RIXS weights 
to microscopic orbital-excitation processes. 
Our calculation supports and further goes beyond the previous 
phenomenological discussion. 

\acknowledgements
It is a great pleasure for the author to thank Dr. Kenji Ishii and Prof. Hiroaki Ikeda 
for invaluable communications. 

\appendix

\section{Hartree-Fock Approximation}
\label{Ap:HF}

Fitting to the first-principles electronic structure of the paramagnetic state, 
one-particle energy levels are determined for the Cu-$d$ orbitals as: 
$\varepsilon_{xy} = -1.84$ eV, $\varepsilon_{xz} = -2.10$ eV, 
$\varepsilon_{yz} = -1.78$ eV, $\varepsilon_{x^2-z^2} = -1.62$ eV, 
$\varepsilon_{3y^2-r^2} = -1.41$ eV at A sites, 
and $\varepsilon_{xy} = -1.84$ eV, $\varepsilon_{yz} = -2.10$ eV, 
$\varepsilon_{xz} = -1.78$ eV, $\varepsilon_{y^2-z^2} = -1.62$ eV, 
$\varepsilon_{3x^2-r^2} = -1.41$ eV at B sites, 
with respect to the Fermi level. 
Here we consider that these values do not reflect 
a realistic level scheme of the local Cu-$d$ orbitals, 
when we perform the HF calculation below. 
Therefore, we modify the one-particle energy levels of the Cu-$d$ orbitals: 
we subtract 2.9 eV from $\varepsilon_{\ell}$ for $d\varepsilon$ orbitals 
and 2.6 eV for $d_{3x^2-r^2}$ and $d_{3y^2-r^2}$ orbitals. 
The orbitals whose one-particle energy is here subtracted 
from should be almost completely filled with electrons, as well known 
from previous studies~\cite{Kadota1967,Liechtenstein1995,Pavarini2008,Deisenhofer2008}. 
This modification allows us to reproduce the electronic structure consistent 
with the observed magnetic ground state within the below HF calculation. 
Hereafter, we redefine $\varepsilon_{\ell}$ by the subtracted one-particle energy. 

\begin{figure}
\begin{center}
\includegraphics[width=90mm]{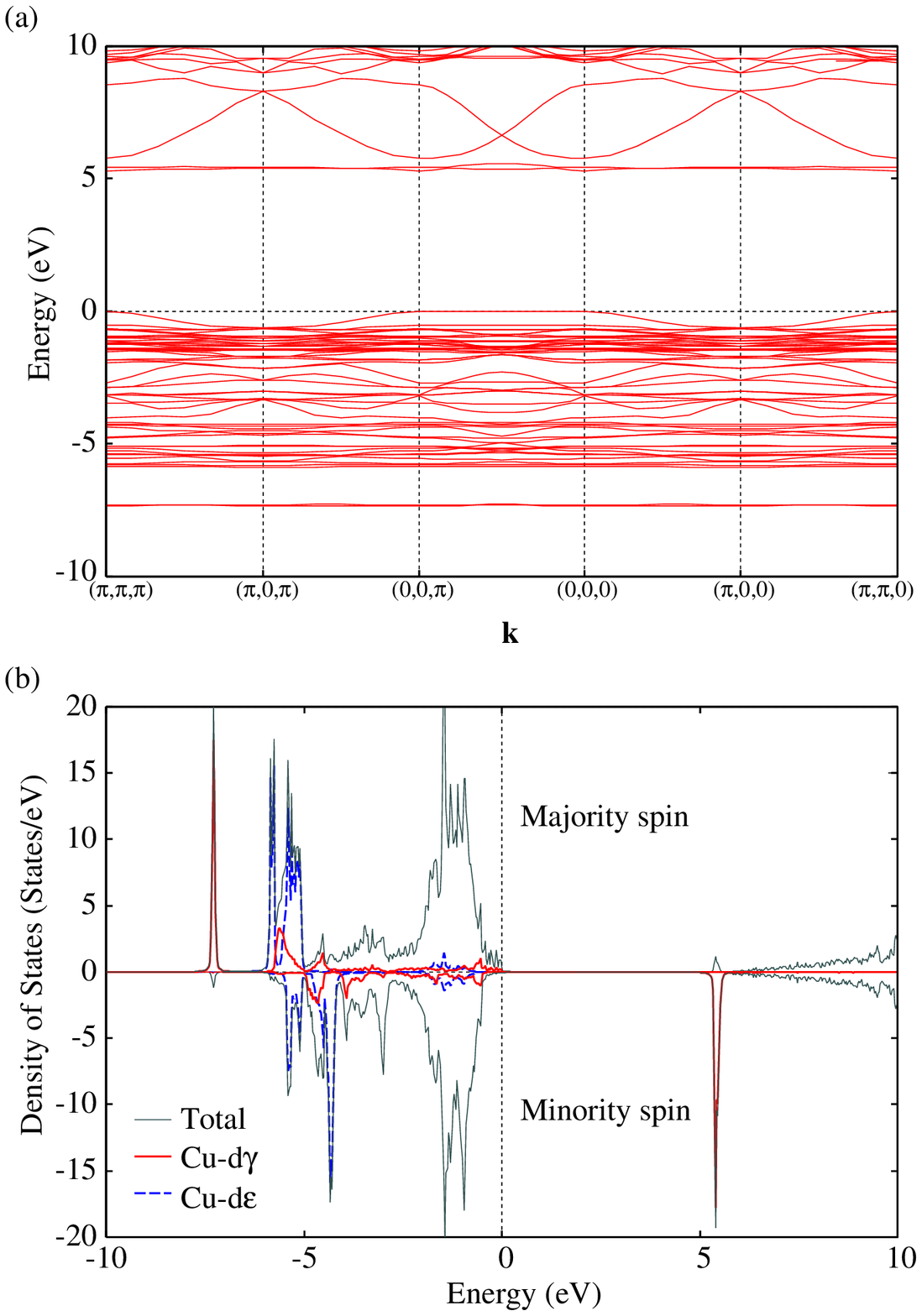}
\end{center}
\caption{
(Color online)
(a) Band structure for the antiferromagnetic ground state calculated 
within the Hartree-Fock approximation. 
(b) Calculated density of states. Thin solid, thick solid and dashed lines represent the total 
density of states, the partial density of states 
of the Cu-$d\gamma$ and Cu-$d\varepsilon$ orbitals, respectively. }
\label{Fig:bandsdos}
\end{figure}

To describe the antiferromagnetic ground state as shown in Fig.~\ref{Fig:lattice}, 
we apply the HF approximation to the tight-binding Hamiltonian $H_{n.f.}$. 
For the Coulomb integrals $I_{\ell_1\ell_2;\ell_3\ell_4}$, we introduce the following notation: 
\begin{eqnarray}
J_{\ell\ell'} & \equiv & I_{\ell\ell';\ell'\ell} \\ 
K_{\ell\ell'} & \equiv & I_{\ell\ell';\ell\ell'}. 
\end{eqnarray}
$J_{\ell\ell'}$ and $K_{\ell\ell'}$ are the so-called direct and exchange integrals, respectively. 
We assume spin polarization is induced only in the $d$ orbitals at Cu site, 
and take mean fields only for the Cu-$d$ electrons. 
The mean-field Hamiltonian for $H_{n.f.}$ is 
\begin{eqnarray}
\label{Eq:HMF}
H_{n.f.}^{MF} &=& \sum_{ii'} \sum_{\ell\ell'} \sum_\sigma 
t_{\ell\ell'}({\mib r}_i-\mib{r}_{i'}) a_{i \ell \sigma}^{\dag} a_{i' \ell' \sigma} 
+ \sum_i^{{\rm t.m.}} \sum_{\ell}^{@\mib{r}_i} 
\biggl[ \frac{J_{\ell\ell}}{2} \langle n_{i\ell} \rangle 
+ \sum_{\ell'(\neq \ell)}^{@\mib{r}_i}  \biggl (J_{\ell\ell'}- \frac{K_{\ell\ell'}}{2}  \biggr) 
\langle n_{i\ell'} \rangle \biggr] n_{i\ell} \nonumber\\
&& - \sum_i^{{\rm t.m.}} \sum_{\ell}^{@\mib{r}_i} 
\biggl[ \frac{J_{\ell\ell}}{2} \langle \mib{m}_{i\ell} \rangle 
+ \sum_{\ell'(\neq \ell)}^{@\mib{r}_i} 
\frac{K_{\ell\ell'}}{2} \langle \mib{m}_{i\ell'} \rangle \biggr] \cdot \mib{m}_{i\ell} 
- \sum_i^{{\rm t.m.}} \sum_{\ell}^{@\mib{r}_i}  \frac{J_{\ell\ell}}{4} 
\biggl( \langle n_{i\ell} \rangle^2 - | \langle \mib{m}_{i\ell} \rangle |^2 \biggr) \nonumber \\
&& - \sum_i^{{\rm t.m.}} \sum_{\ell \neq \ell'}^{@\mib{r}_i} 
\frac{J_{\ell\ell'}}{2} \langle n_{i\ell} \rangle \langle n_{i\ell'} \rangle 
+ \sum_i^{{\rm t.m.}} \sum_{\ell \neq \ell'}^{@\mib{r}_i}  \frac{K_{\ell\ell'}}{4} 
\biggl( \langle n_{i\ell} \rangle \langle n_{i\ell'} \rangle 
+ \langle \mib{m}_{i\ell} \rangle \cdot \langle \mib{m}_{i\ell'} \rangle \biggr), 
\end{eqnarray}
where 
\begin{eqnarray}
n_{i\ell} &=& \sum_{\sigma} d_{i\ell\sigma}^{\dag} d_{i\ell\sigma} \\
\mib{m}_{i\ell} &=& \sum_{\sigma\sigma'} d_{i\ell\sigma}^{\dag} 
\mib{\sigma}_{\sigma\sigma'} d_{i\ell\sigma'}, 
\end{eqnarray}
using the Pauli matrix vector $\mib{\sigma}$. 
Within the HF theory, we should consider that the one-particle energy $\varepsilon_{\ell}$ 
is already including the following energy shift from the bare one, 
\begin{equation}
\Delta^{HF}_{\ell} \equiv \frac{J_{\ell\ell}}{2} \langle n_{i\ell} \rangle 
+ \sum_{\ell'(\neq \ell)}^{@\mib{r}_i}  \biggl (J_{\ell\ell'}- \frac{K_{\ell\ell'}}{2}  \biggr) 
\langle n_{i\ell'} \rangle, 
\label{Eq:Delta}
\end{equation}
due to the electron-electron Coulomb interaction at transition-metal site $\mib{r}_i$. 
Therefore, before determining the magnetic ground state, 
we need evaluate the bare one-particle energy by 
$\varepsilon_{\ell}^{(0)} \equiv \varepsilon_{\ell} - \Delta^{HF}_{\ell}$, 
where $\Delta^{HF}_{\ell}$ is evaluated from the expectation values 
of particle numbers $\langle n_{i\ell} \rangle$'s in the paramagnetic state 
using eq.~(\ref{Eq:Delta}). 
For the Coulomb integrals given in \S~\ref{Sc:Formulation}, 
the obtained values of $\varepsilon_{\ell}^{(0)}$ are as follows: 
$\varepsilon_{xy}^{(0)} = -86.5$ eV, $\varepsilon_{xz}^{(0)} = -86.1$ eV, 
$\varepsilon_{yz}^{(0)} = -86.5$ eV, $\varepsilon_{x^2-z^2}^{(0)} = -85.7$ eV, 
$\varepsilon_{3y^2-r^2}^{(0)} = -86.2$ eV at A sites, 
and $\varepsilon_{xy}^{(0)} = -86.5$ eV, $\varepsilon_{yz}^{(0)} = -86.1$ eV, 
$\varepsilon_{xz}^{(0)} = -86.5$ eV, $\varepsilon_{y^2-z^2}^{(0)} = -85.7$ eV, 
$\varepsilon_{3x^2-r^2}^{(0)} = -86.2$ eV at B sites, with respect to the Fermi level. 
We consider that these values of $\varepsilon_{\ell}^{(0)}$ may reflect a realistic level 
scheme of the local Cu-$d$ orbitals: 
$\varepsilon_{x^2-z^2}$ and $\varepsilon_{y^2-z^2}$ are the highest level 
among the five local Cu-$d$ levels, as several studies suggest~\cite{Deisenhofer2008}. 
Maintaining these values of $\varepsilon_{\ell}^{(0)}$, we determine the mean-fields 
$\langle n_{i\ell} \rangle$ and $\langle \mib{m}_{i\ell} \rangle$ self-consistently. 
As a result, we obtain 104 diagonalized energy bands 
($E_j(\mib{k})$, $1 \leq j \leq 104$) for the antiferromagnetic ground state. 
The Cu-$d_{x^2-z^2}$ and Cu-$d_{y^2-z^2}$ orbitals are nearly half-filled, 
while the other Cu-$d$ orbitals are almost fully filled. 
Figure~\ref{Fig:bandsdos} shows the obtained electronic structure 
and the density of states for the orbital-ordered antiferromagnetic ground state. 
Flat bands around 5 eV and $-7$ eV with respect to the Fermi energy correspond 
to the upper and lower Hubbard bands, respectively. 
In the region between the upper and lower Hubbard bands, 
Cu-$d$ electronic states hybridize with F-$p$ states. 
Therefore our electronic structure suggests that KCuF$_3$ 
lies in the charge-transfer regime, rather than in the Mott-Hubbard regime. 
The bands between the upper and lower Hubbard bands seem rather flat, 
compared with the cases of copper oxides~\cite{Nomura2004,Nomura2005}, 
and this is a reason why the RIXS spectra do not show notable momentum 
dependence in KCuF$_3$. 

\end{document}